\documentclass[12 pt]{iopart}

\usepackage{graphicx}
\usepackage{iopams}

\newcommand{\be}[1]{\begin{equation} \label{#1}}
\newcommand{\ee}{\end{equation}}
\newcommand{\bem}[1]{\begin{eqnarray} \label{#1}} \newcommand{\eem}{\end{eqnarray}}
\newcommand{\bems}[1]{\numparts \begin{eqnarray} \label{#1}} \newcommand{\eems}{\end{eqnarray} \endnumparts}
\newcommand{\ce}{\\ && \nonumber \quad} \newcommand{\nems}[1]{\\ \label{#1}} \newcommand {\ket}[1]{\vert \, #1\rangle}
\newcommand {\bra}[1]{\langle #1 \, |}
\newcommand {\braket}[2]{\langle #1 \, | \, #2 \rangle}
\def\A{{a^{\dagger}}}
\def\a{a}
\def\B{{b}^{\dagger}}
\def\b{b}
\def\C{{c}^{\dagger}}
\def\c{c}
\def\D{{d}^{\dagger}}
\def\d{d}
\def\I{ \scriptsize{\textrm{Im}}}	
\begin{document}

\title[Verifying a Quantum Superposition in a Micro-optomechanical System]{Creating and Verifying a Quantum Superposition in a Micro-optomechanical System}
\author{Dustin Kleckner$^{1, 2}$, Igor Pikovski$^{1, 3, 4}$, Evan Jeffrey$^3$, Luuk Ament$^5$, Eric Eliel$^3$, Jeroen van den Brink$^{5, 6}$, Dirk Bouwmeester$^{2,3}$}
\address{
	$^1$These authors contributed equally to this work.\\
	$^2$Physics Department, University of California, Santa Barbara\\
	$^3$Huygens Laboratory, Universiteit Leiden\\
	$^4$Fachbereich Physik, Freie Universit\"{a}t Berlin\\
	$^5$Institute-Lorentz for Theoretical Physics, Universiteit Leiden \\
	$^6$Institute for Molecules and Materials, Radboud Universiteit Nijmegen\\
}
\ead{\mailto{dkleckner@physics.ucsb.edu}, \mailto{pikovski@molphys.leidenuniv.edu}}
\begin{abstract}
Micro-optomechanical systems are central to a number of recent proposals for realizing quantum mechanical effects in relatively massive systems.
Here we focus on a particular class of experiments which aim to demonstrate massive quantum superpositions, although the obtained results should be generalizable to similar experiments.
We analyze in detail the effects of finite temperature on the interpretation of the experiment, and obtain a lower bound on the degree of non-classicality of the cantilever.
Although it is possible to measure the quantum decoherence time when starting from finite temperature, an unambiguous demonstration of a quantum superposition requires the mechanical resonator to be in or near the ground state.
This can be achieved by optical cooling of the fundamental mode, which also provides a method to measure the mean phonon number in that mode.
We also calculate the rate of environmentally induced decoherence and estimate the timescale for gravitational collapse mechanisms as proposed by Penrose and Diosi. 
In view of recent experimental advances, practical considerations for the realization of the described experiment are discussed.
\end{abstract}
\maketitle

\section{Introduction}
Micro-optomechanical systems have recently attracted significant interest as a potential architecture for observing quantum mechanical effects on scales many orders of magnitude more massive than previous experiments.
Proposals include entangling states of mechanical resonators to each other~\cite{Mancini2003EPJD, Pinard2005EPL, Vitali2007JPA} or cavity fields~\cite{Vitali2007, Paternosto2007PRL}, the creation of entangled photon pairs~\cite{Giovannetti2001}, ground state optical feedback cooling of the fundamental vibrational mode~\cite{Courty2001, WilsonRae2007, Marquardt2007PRL, Bhattacharya2007PRL, Bhattacharya2008PRA}, observation of discrete quantum jumps~\cite{Thompson2008}, quantum state transfer~\cite{Zhang2003PRA}  and the creation of massive quantum superpositions or so-called ``Schr\"odinger's cat'' states~\cite{Bose1997PRA, Bose1999PRA, Marshall2003PRL}.
Here we focus on the latter class of experiments, in particular the one as described in Marshall et al.~\cite{Marshall2003PRL}.

\begin{figure}[b]
\begin{center}
	\includegraphics{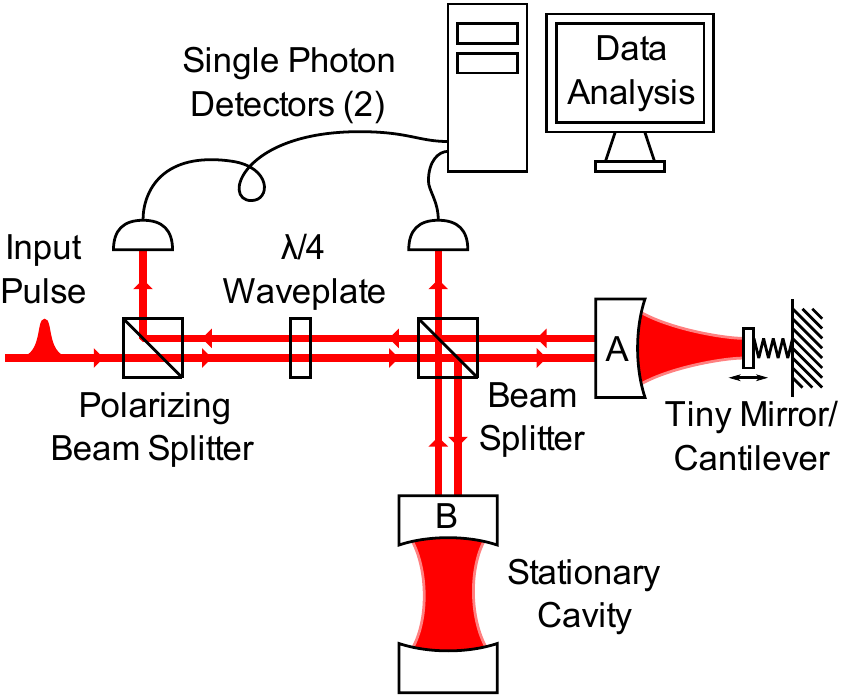}
\end{center}
\caption{
	A diagram of the experimental setup.
	An input pulse is split between the two arms of a Michelson interferometer, labeled A and B, both of which contain high finesse cavities.
	One end of the cavity in arm A is a tiny end mirror on a micromechanical cantilever, whose motion is affected by the radiation pressure of light in the cavity.
	Each output port of the interferometer is monitored by a single photon detector, and results are analyzed by a computer to calculate the interference visibility.
}
\label{fig-setup}
\end{figure}

The heart of this experiment is a Michelson interferometer with high finesse optical cavities in each of its arms (\fref{fig-setup}).
In one arm the traditional end mirror is replaced with a tiny mirror on a micromechanical cantilever, hereafter referred to as the ``cantilever''.
Under the right conditions, the radiation pressure of a single photon in this arm of the experiment will be enough to excite the cantilever into a distinguishable quantum state.
A single photon incident on the 50-50 beam splitter will form an optical superposition of being in either of the two arms; the coupling between the photon and the cantilever will then entangle their states, putting the cantilever into a superposition as well.
If the photon leaves the interferometer with the cantilever in a distinguishable state, an outside observer could in principle determine which arm the photon took, and so the interference visibility is destroyed.
After a full mechanical period of the cantilever, however, it returns to its original position: if the photon leaves the interferometer at this time, the interference visibility should return provided the cantilever was able to remain in a quantum superposition in the intermediate period.
Alternatively, if the state of the cantilever collapses during this period due to environmentally induced decoherence, measurement by an outside observer or perhaps an exotic mechanism (e.g. \cite{Karolyhazy1966, Penrose1996, Diosi1989PRA}), the visibility will not return.
In this sense the interference revival constitutes evidence that the cantilever was able to exist in a quantum superposition, and a measurement of its magnitude constitutes a measurement of the quantum decoherence in this time interval.
In a real experiment, however, one must be careful about drawing conclusions from the visibility dynamics as similar results can be obtained from a fully classical argument.

In this work we address the issue of classicality by first calculating the quantum dynamics of the system for both a pure state and a thermal density matrix (\sref{section-quantum}).
We also caclulate the Wigner function of the system as a method of determining the transition from the quantum to classical regime (\sref{section-wigner}).
Finally we discuss quantum decoherence mechanisms (\sref{section-decoherence}) and prospects for realization in view of recent experimental results (\sref{section-experiment}).

\section{Quantum Mechanical Description}
\label{section-quantum}
A more detailed analysis of the system begins with the Hamiltonian, given by Law ~\cite{Law1995PRA}:
\be{eq-Hamiltonian}
H = \hbar \omega_a \left[ \A \a + \B \b \right] + \hbar \omega_c \left[\C \c - \kappa \A \a \left(\c + \C\right)\right],
\ee
where $\omega_a$ is the frequency of the optical field, $\A$/$\B$ and $\a$/$\b$ are the the photon creation and annihilation operators for photons the arms A and B of the interferometer, $\omega_c$ is the mechanical frequency of the cantilever and $\C$ and $\c$ are the phonon creation and annihilation operators for its fundamental vibrational mode.
The dimensionless opto-mechanical coupling constant $\kappa$ is defined as:
\bems{eq-Kappa}
\kappa &=& \frac{\omega_a}{L \omega_c} \sqrt{\frac{\hbar}{2 m \omega_c}}
 \nems{eq-Kappa-2}
&=& \frac{\sqrt{2} N x_0}{\lambda},
\eems
where $m$ is the mass of the cantilever, $L$ is the length of the optical cavity, $N$ is the number of cavity round trips per mechanical period, $\lambda$ is the optical wavelength, and $x_0 = \sqrt{\frac{\hbar}{m \omega_c}}$ is the size of the ground state wavepacket for the cantilever.
The Hamiltonian treats the mechanical resonator as completely linear, which should be a valid assumption.
Non-linearities have not been observed in experiments conducted on similar systems, which is expected given that the typical vibration amplitudes are many orders of magnitude smaller than the dimensions of the resonator.
From this we can derive the unitary evolution operator ~\cite{Bose1997PRA}:
\vbox{
\bem{eq-unitary}
U(t)&=&\exp\Big[-i \omega_a t \left(\A\a + \B\b\right) - i \left(\kappa \A \a\right)^2 \left(\omega_c t - \sin \omega_c t\right) \Big] \times
\ce \exp\Big[ \kappa \A \a \left[ \left(1-e^{-i \omega_c t}\right) \C - \left(1-e^{i \omega_c t}\right)\c \right] \Big]  \exp\Big[- i \omega_c \C \c t \Big].
\eem
}
\subsection{Coherent State}
If we consider a cantilever initially in a coherent state with complex amplitude $\beta$, the total initial state is given by
$\ket{\Psi(0)} = \frac{1}{\sqrt{2}} \left( \ket{0,1}_{n_a, n_b}  + \ket{1,0}_{n_a, n_b} \right) \otimes \ket{\beta}_c$.
Under the action of the unitary operator eqn.~\eref{eq-unitary} this unentangled state evolves to:
\bems{eq-state-1}
	\ket{\Psi(t)}&=&\frac{1}{\sqrt{2}} e^{-i \omega_a t}
	\Big( \ket{0, 1} \otimes \ket{\beta e^{-i \omega_c t}} +
	\ce e^{i \kappa^2 (\omega_c t - sin(\omega_c t)) +
	i \kappa \I [ \beta (1- e^{-i \omega_c t}) ] }
	\ket{1, 0} \otimes \ket{\kappa (1- e^{-i \omega_c t}) + \beta e^{-i \omega_c t}} \Big)
\nems {eq-state-2}
	&=&\frac{1}{\sqrt{2}} e^{-i \omega_a t}
	\Big( \ket{0, 1} \otimes \ket{\Phi_0(t)} +
	\ce e^{i \kappa^2 (\omega_c t - sin(\omega_c t)) -
	i \I [ \Phi_0(t) \Phi_1(t)^*]} \ket{1, 0} \otimes \ket{\Phi_1(t)} \Big).	\eems
Because the cantilever is only displaced if the photon is in arm A, the state of the photon and the state of the cantilever become entangled.
The cantilever then enters a superposition of two different coherent states, with time dependent amplitude $\Phi_0(t)$ when no photon is present and $\Phi_1(t)$ if there is a photon.
After half a mechanical period, the spatial distance between the two cantilever states $\ket{\Phi_0}$ and $\ket{\Phi_1}$ is given by $\Delta x = \sqrt{8} \kappa x_0$, and the two cantilever states have the lowest overlap, $|\braket{\Phi_0}{\Phi_1}| = e^{-2 \kappa^2}$.
After a full mechanical period $\ket{\Phi_0}$ and $\ket{\Phi_1}$ are identical again, and so the photon and cantilever are disentangled.
For a proper demonstration of a superposition, we require the overlap between the states to be relatively small during part of the experiment, implying $\kappa \gtrsim 1/\sqrt{2}$.
This is equivalent to stipulating that a measurement of the cantilever state alone is sufficient to determine which path a photon took with a reasonable fidelity.
As will be discussed in \sref{section-experiment}, obtaining this large a value of $\kappa$ poses the most significant barrier to experimental realization.

In practice, the actual quantity measured is the interferometric visibility $v(t)$ as seen by the two single photon detectors.
This visibility is given by twice the absolute value of the off-diagonal elements of the reduced photon density matrix, which in this case is the overlap between the two cantilever states:
\be{eq-visibility}
v(t) = e^{-\kappa^2 (1-\cos(\omega_c t))}.
\ee
\begin{figure}[!htp]
	\begin{center}
		\includegraphics{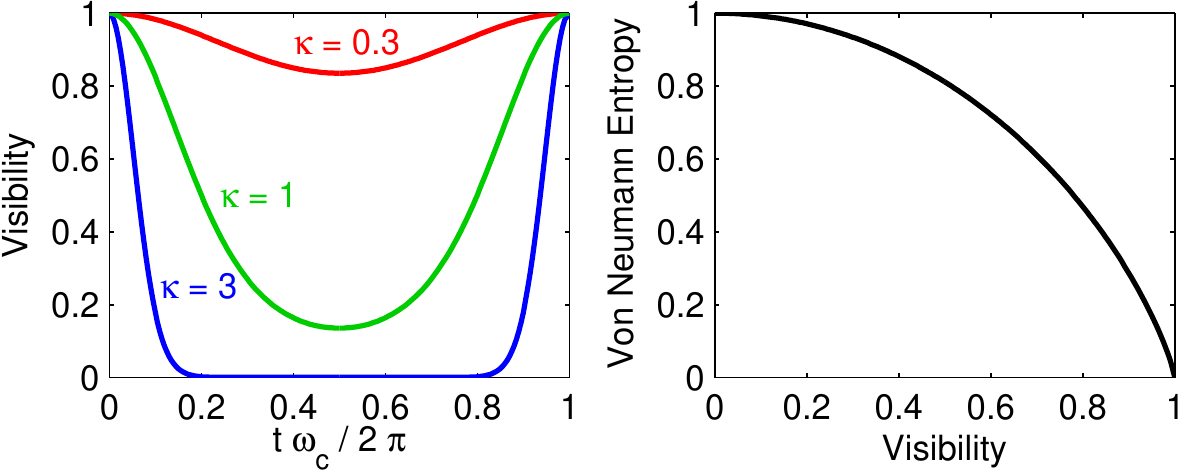}
	\end{center}
	\caption{Left: The Visibility $v(t)$ as a function of time for different values of the opto-mechanical coupling constant, $\kappa$.
			Right: The 	Von Neumann entropy $S(t)$ versus the visibility, $v(t)$.}
	 \label{fig:Vis}
\end{figure}
It exhibits a periodic behavior characterized by a suppression of the interference visibility after half a mechanical period and a revival of perfect visibility after a full period (\fref{fig:Vis}) provided there is no decoherence in the state of the cantilever.
The visibility can be mapped directly to the entanglement between the photon and the cantilever. For a pure bipartite state, we can express the entanglement as the von Neumann entropy of the photon $S(t)$ in terms of the visibility $v(t)$ (\fref{fig:Vis}):

\bems{eq-Entropy-1}
	S(t)&=&-\textrm{Tr}_{\textrm{ph}} \left( \rho_{\textrm{ph}} \log_2 \rho_{\textrm{ph}} \right)
\nems{eq-Entropy-2}
	&=&1+\frac{v(t)}{2} \log_2 \left( \frac{1-v(t)}{1+v(t)} \right) - \frac{1}{2} \log_2 \left( 1-v(t)^2 \right),
\eems
where $\rho_{\textrm{ph}}$ is the reduced density matrix for the photon.
Since for a pure bipartite system a high Von Neumann entropy of one subsystem corresponds to high entanglement between the two subsystems, we conclude that when the initial state is pure, the visibility alone is a good measure for the non-classical behavior of the cantilever.
This is true even in the presence of an arbitrary decoherence mechanism, which will destroy the quantum nature of the system and thus produce a corresponding loss of interference visibility.

\subsection{The cantilever at finite temperatures}

At finite temperatures the exact wavefunction of the cantilever is unknown, so the state is instead described by a density matrix:
\be{eq-thermal-state}
\rho_c(0) = \frac{\sum_n e^{-E_n/k_B T } \ket{n} \bra{n} }{ \sum_n e^{-E_n/k_B T } }  = \frac{1}{\pi \bar{n}} \int d^2\beta e^{-|\beta|^2 / \bar{n} } \ket{\beta} \bra{\beta},
\ee
where $\bar{n} = 1/ ( e^{\hbar \omega_c / k_B T} - 1) $ is the average thermal occupation number of the cantilever's center of mass mode, $\ket{n} $ are energy eigenstates and $\ket{\beta} $ coherent states of the cantilever.
Here we only consider the effects of a thermally excited initial state, i.e. for a cantilever with no dissipation ($Q \to \infty$).
The effects of dissipation and resulting decoherence are discussed in \sref{section-decoherence}.

The evolution of eqn.~\eref{eq-thermal-state} under the action of eqn.~\eref{eq-unitary} yields the visibility:
\be{eq-visibilityTH}
v(t) = e^{-\kappa^2 (2 \bar{n} +1)  (1-\cos (\omega_c t))}.
\ee
At finite temperatures the density matrix represents an average over coherent states with different phases which destroys the interference visibility.
Although there is also a phase shift from the entanglement as discussed earlier, in principle this shift is known and repeatable, while the same is not true for the thermal state.
A good indicator that the visibility no longer captures the quantum behavior is that it becomes independent of $\hbar$ if the initial temperature of the cantilever is high \cite{Bernad2006PRL}.
This can be seen most easily by noting that in the limit $k_b T \gg \hbar \omega_c$, the mean phonon number is given by $\bar{n} \approx k_b T/\hbar \omega_c - 1/2$.  Thus the visibility eqn.~\eref{eq-visibilityTH} can be rewritten as:
\be{eq-visibility-highT}
v(t) \approx e^{-\frac{k_b T}{m \omega_c^2} \left(\frac{2 N}{\lambda}\right)^2 \left(1 - \cos (\omega_c t)\right)}.
\ee
This is the classically expected result, which differs primarily from the quantum result in that the visibility is always one at zero temperature because the distinguishability of the cantilever state is irrelevant.
At higher temperatures it is difficult to determine when the cantilever was in a superposition state. 
Because the experiment requires averaging over many runs, the quantum distinguishability is masked by the unknown classical phase shifts.

However, after a full mechanical period the net phase shift from any initial state goes to zero and so full visibility should still return in a narrow window whose width scales like $\bar{n}^{-1/2}$.
This leaves open the possibility for measuring quantum collapse mechanisms at higher temperatures if one assumes that the cantilever was in a superposition state.
Provided that the opto-mechanical coupling strength $\kappa$ is relatively well known (e.g., by independently measuring $m$, $\omega_c$, $L$, etc.) and the instantaneous quantum state of the cantilever is regarded as some random coherent state (as should be the case for the weakly mechanically damped systems discussed here) it can be easily determined when a superposition should have been created.

Although eqn.~\eref{eq-visibility-highT} suggests the visibility should always return in the classical case, we note that this can only be true if both the optical \emph{and} mechanical modes are behaving classically.
On the other hand, if we regard only the optical field as quantum we should always expect no interference visibility because the classical cantilever would measure which path the photon took.
Thus the return of visibility at higher temperatures can be used to strongly imply the existence of a quantum superposition when $\kappa \gtrsim 1/\sqrt{2}$, even though the superposition can not be directly measured by the visibility loss at $t \sim \pi \omega_c$\footnote {The presence of a ``loop hole'' in such a demonstration could be regarded as analogous to experimental tests of Bell's inequalities, where even though it is generally regarded that quantum mechanics has been adequately demonstrated, an unambiguous proof has remained elusive.
In our case, the loop hole is caused by the unknown intermediate state caused by finite temperature. 
Even though a weakly damped system should produce something that is very nearly a coherent state at any given instance of time, there is no way to directly show the cantilever is in this state.
}.

Nevertheless, an unambiguous demonstration can be provided if the temperature is low enough such that the visibility loss due to quantum distinguishability is still resolvable.
At finite cantilever temperatures the interferometric visibility becomes a bad measure for the non-classicality of the mirror.
This can be easily seen by the the relation between the von Neumann entropy and the visibility, eqn.~\eref{eq-Entropy-2}.
It is valid at arbitrary temperatures, but at $T>0$ the system is in a mixed state and the entropy is only an upper bound for the entanglement of formation \cite{Nielsen2001Book}.
One thus needs to analyze the non-classicality of the cantilever state by other means.
In the next section we use the integrated negativity of the Wigner function~\cite{Kenfack2004JOptB} to quantify the non-classicality of the cantilever with respect to temperature.

\section{The Wigner Function and the Classical Limit}
\label{section-wigner}
\begin{figure}[!htp]
	\begin{center}
		\includegraphics[width=6.0 in]{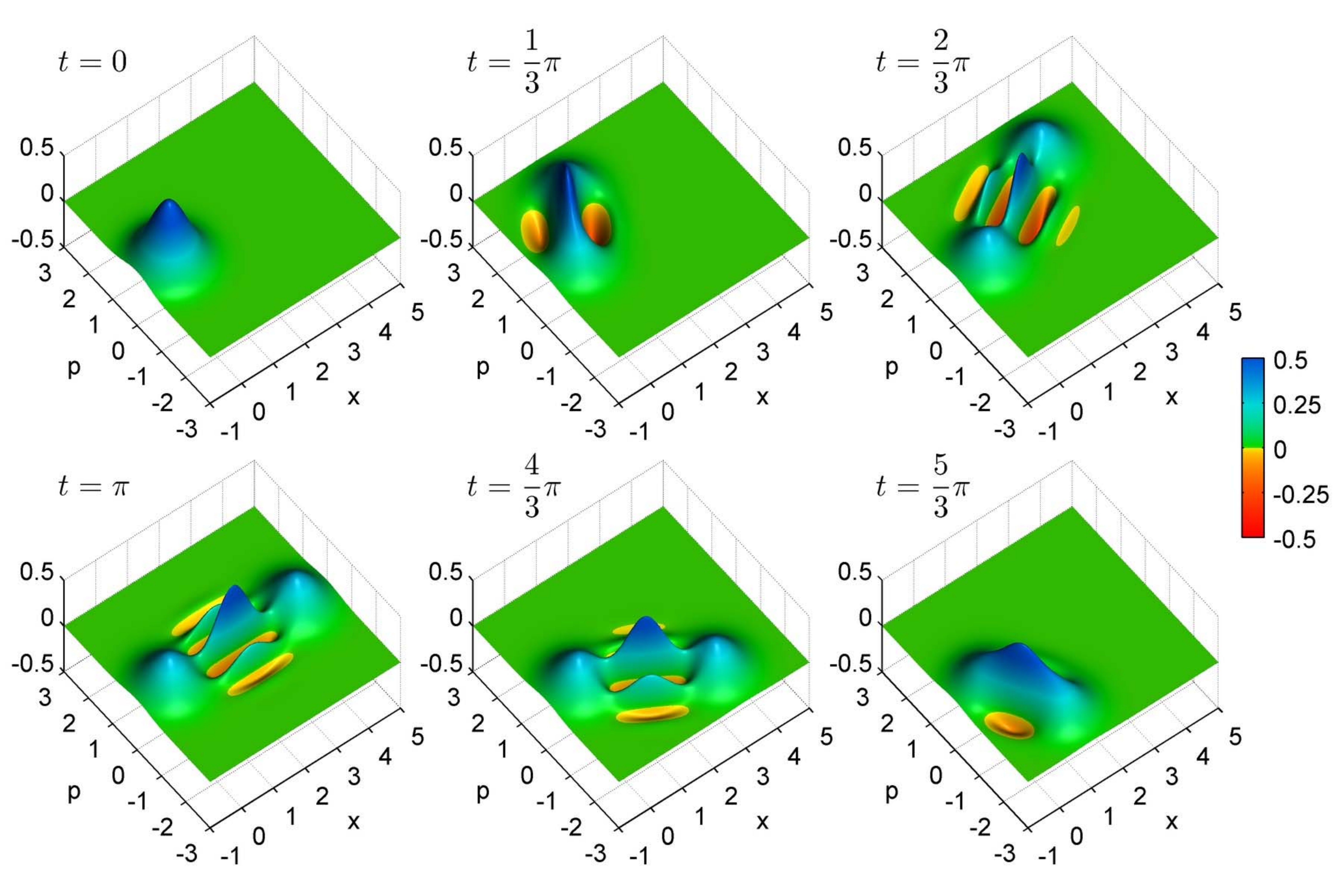}
	\end{center}
	\caption{
		The time evolution of the cantilever's projected Wigner function for $\beta = 0$, $\kappa =2$ and $\hbar = \omega_c = m = 1$.
		Regions where the Wigner function is negative, shown in yellow and red, have no classical analogue.
	}
\label{fig:Wigner-Ground}
\end{figure}

To study transitions between the quantum and the classical regimes, it is often convenient to refer to quasi-probability distributions, with which quantum mechanics can be formulated in the common classical phase space.
One such distribution was proposed in 1932  by Wigner \cite{Wigner1932PhysRev} and can be obtained from the density matrix $\rho$:
\be{Wig}
W(x,p) = \frac{1}{\pi \hbar} \int_{- \infty}^{+\infty} dy \bra{x-y} \rho \ket{x+y} e^{2ipy/\hbar}.
\ee
It is well known that in the classical limit $\hbar \rightarrow 0 $ the Wigner function tends to a classical probability distribution describing a microstate in phase space \cite{Hillery1984PhR}.
This can most easily be seen in the case of a single particle moving in a potential $V(x)$.
The time evolution of the Wigner function for this closed system is described by the quantum Liouville equation \cite{Wigner1932PhysRev, Schleich2001book}

\bem{eq-WignerLiouville}
&&\big( \frac{\partial}{\partial t} + \frac{p}{m} \frac{\partial}{\partial x}  - \frac{dV(x)}{dx} \frac{\partial}{\partial p} \big) \, W(x, p,t) =
\ce\sum_{k=1}^{\infty} \hbar^{2k} \frac{(-1)^k}{4^k (2k+1)!} \frac{d^{2k+1}V(x)}{dx^{2k+1}}
\frac{\partial^{2k+1}}{\partial p^{2k+1}} W(x,p,t).
\eem
For $\hbar \rightarrow 0 $, the right hand side goes to 0, as long as no derivatives diverge.
In this limit the Wigner function $W(x,p,t)$ thus evolves according to the classical Liouville equation.
However, the quantum nature of $W(x,p,t)$ is also contained in its initial conditions.
In fact, in the special case of a harmonic potential, all non-classical behavior is encoded in the initial conditions of the Wigner function only since the right hand side of eqn.~\eref{eq-WignerLiouville} is always 0.
But for $\hbar \rightarrow 0$ also the initial conditions become classical and $W(x,p,t)$ can be fully identified with some classical probability density.

If, on the other hand, the Wigner function is negative then no classical interpretation is possible, making it a useful tool to indicate the non-classicality of an arbitrary state.
It is thus convenient to quantify the total negativity of the Wigner function \cite{Kenfack2004JOptB}:
\bem{eq-Negativity}
 && N =  \int_{- \infty}^{+\infty}\!\!  dx \int_{- \infty}^{+\infty}\!\!  dp \Big\{ |W(x,p)| - W(x,p) \Big\}
	\ce  =  \int\!\!  dx \int\!\!  dp  \, |W(x,p)| - 1.
\eem

For the experiment at hand, we compute the cantilever's Wigner function for dimensionless $x$ and $p$, with the photon projected into the superposition state $\ket{0,1} + e^{i \theta} \ket{1,0}$ to avoid destroying the quantum state of the cantilever to which it is entangled.
This projection is equivalent to detecting a single photon at one output of the interferometer, where the phase term in the projection accounts for path length differences in the arms.
Generally speaking, varying $\theta$ shifts the interference peaks but does not modify the Wigner function in a significant way; hereafter we will set it to 0.
The resulting Wigner function of the cantilever indeed shows that the system periodically exists in a highly non-classical state (\fref{fig:Wigner-Ground}).

\begin{figure}[!htp]
	\begin{center}
		\includegraphics[width=6.0 in]{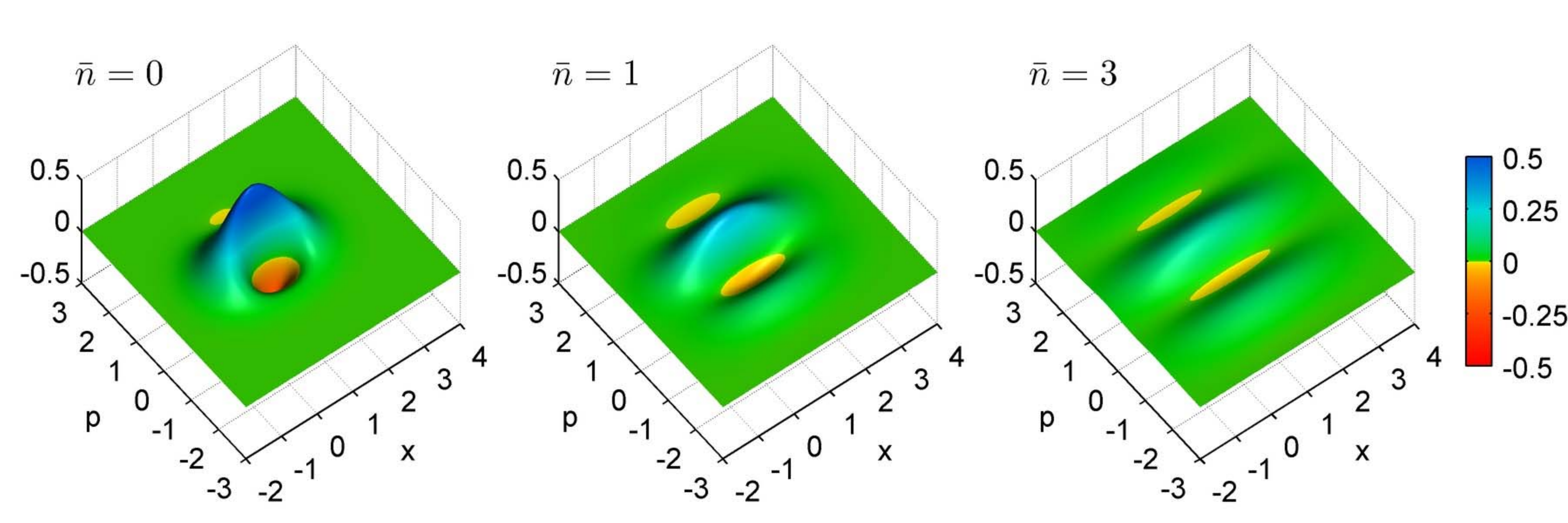}
	\end{center}
	\caption{
		The thermally averaged projected Wigner function of the cantilever at time $t=\pi$ for $\kappa = 1/\sqrt{2}$ and different mean thermal phonon numbers, $\bar{n}$ .
		($\hbar = \omega_c = m = 1$)
		The negative regions of the Wigner function, shown in yellow and red, can be seen to quickly wash out with increasing temperature.
	}
	\label{fig:Wigner-Thermal}
\end{figure}

\begin{figure}[!htp]
	\begin{center}
		\includegraphics{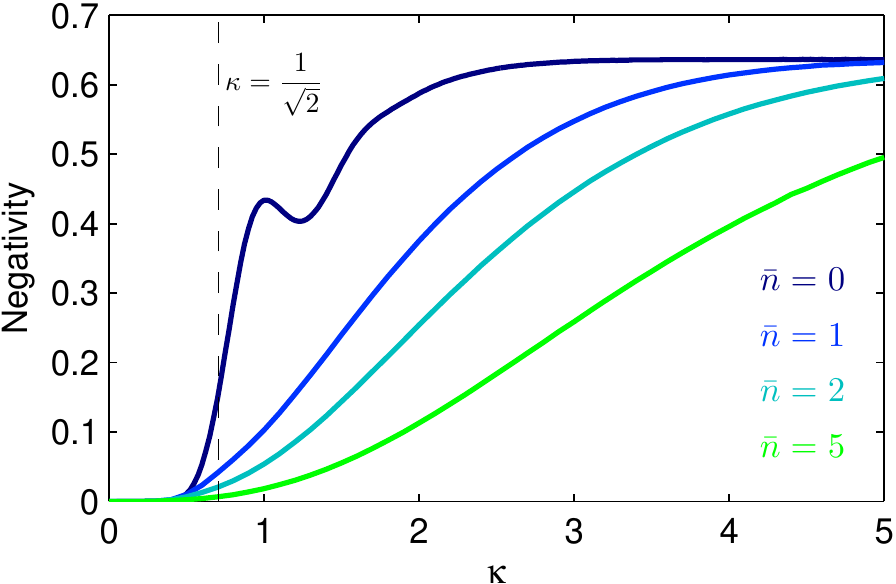}
	\end{center}
	\caption{
		Negativity of the projected cantilever state as a function of coupling constant $\kappa$ for several different mean phonon numbers, $\bar{n}$.
		The oscillations present when $\bar{n}=0$ are due to a phase shift in the interference terms, which are washed out at higher temperatures.
	}
	\label{fig:Negativity}
\end{figure}

A calculation of the thermally averaged Wigner function shows that the non-classical features are quickly washed out with increasing initial temperature (\fref{fig:Wigner-Thermal}).
However, as long as part of the Wigner function is negative, the cantilever is clearly in a non-classical superposition state.
The negativity of the Wigner function at half a mechanical round trip decreases rapidly with $\bar{n} $ and is also dependent on $\kappa$ (\fref{fig:Negativity}).
In practice, this implies that $\bar{n}$ must of order 1 for $\kappa \approx 1$, with somewhat higher values being tolerable for higher $\kappa$.
This analysis confirms our earlier assertion that direct proof of a superposition requires low mean phonon number.

Finally, we mention that it is also possible to demonstrate the non-classical nature of a mechanical resonator by calculating a measure of entanglement~\cite{Paternosto2007PRL}. 
For example, in a related experiment in which two micromechanical systems are coupled to one another with a light field, the entanglement is lost at higher temperatures~\cite{Vitali2007JPA, Vitali2007}
(the larger temperature bound obtained is due to a large amplitude coherent state in the optical mode).

\section{Decoherence}
\label{section-decoherence}
In addition to ``classical'' phase scrambling caused by the initial thermal motion of the cantilever as discussed above, there are other effects which cause ``quantum'' decoherence of the cantilever.
The signature of this type of decoherence is a reduction of the visibility's revival peak -- this is caused by information loss during a single experimental run.
This is different from the previously discussed effect which is a narrowing of the visibility revival peaks caused by averaging of states in a thermal mixture, where no information is lost.
Thus, to be able to detect a signature of a macroscopic superposition, the timescale on which decoherence occurs should be larger than a single mechanical period.

\subsection{Environmentally Induced Decoherence}
Environmentally induced decoherence is due to the coupling of the system to a finite temperature bath, and results in a finite lifetime for the quantum superposition of the cantilever.
Decoherence happens when the thermal bath measures the state of the cantilever while the photon is in the cavity, introducing a phase shift that can not be compensated for even in principle.
To find the time scale for this mechanism we need to solve the open quantum representation of the system.
This is generally done by coupling the cantilever to an infinite bath of harmonic oscillators and integrating out the environmental degrees of freedom.
In doing so, one obtains a time-local master equation for the density matrix of the system incorporating the influence of the environment.

We start with the Hamiltonian:
\be{eq-OpenHamilonian}
H = H_{sys} + H_{bath} + H_{int},
\ee
where:
\bem{eq-OpenHamilonian2}
 & & \quad H_{sys}  =  \hbar \omega_a \left[ \A \a + \B \b \right] + \hbar \omega_c \left[\C \c - \kappa \A \a \left(\c + \C\right)\right]
	\ce H_{bath}  = \sum_i \hbar \omega_i \D_i \d_i
	\ce H_{int}	= (\c + \C) \sum_i \lambda_i (\d_i + \D_i ).
\eem
Here $\D_i$ ($\d_i $) are the creation (annihilation) operators of the bath modes, $\omega_i$ is the frequency of each mode and $\lambda_i $ are coupling constants.
Using the Feynman-Vernon influence functional method \cite{Feynman1963} we can eliminate the bath degrees of freedom.
When the thermal energy of the bath sets the highest energy scale we can use the Born-Markov approximation to obtain a master equation for the density matrix of our system \cite{Caldeira1983PhysA}:
\be{eq-Master}
	\dot{\rho}(t) = \frac{1}{i\hbar} \left[ \tilde{H}_{sys},\rho(t) \right] -\frac{i \gamma}{\hbar} \left[ x,\left\{ p,\rho(t) \right\} \right] - \frac{D}{\hbar^2} \left[ x,\left[ x,\rho(t) \right] \right],
\ee
where $\tilde{H}_{sys}$ is the system Hamiltonian in eqn.~\eref{eq-Hamiltonian}, renormalized by the interaction of the cantilever with the bath.
$\gamma = \omega_c/ Q $ is the damping coefficient as determined from the mechanical Q~factor and $D=2m\gamma k_B T_b $ is the diffusion coefficient where $T_b$ is the temperature of the bath.
The first term on the right hand side of~\eref{eq-Master} is the unitary part of the evolution with a renormalized frequency.
The other terms are due to the interaction with the environment only and incorporate the dissipation and diffusion of the cantilever.
The equation is valid in the Markovian regime when memory effects in the bath can be neglected; this is satisfied when the coupling to the bath is weak ($Q \gg 1$) and the thermal energy is much higher than the phonon energy ($k_B T_b \gg \hbar \omega_c$).
Both conditions are easily satisfied for realistic devices.
\begin{figure}[!htp]
	\begin{center}
		\includegraphics[width=6.0 in]{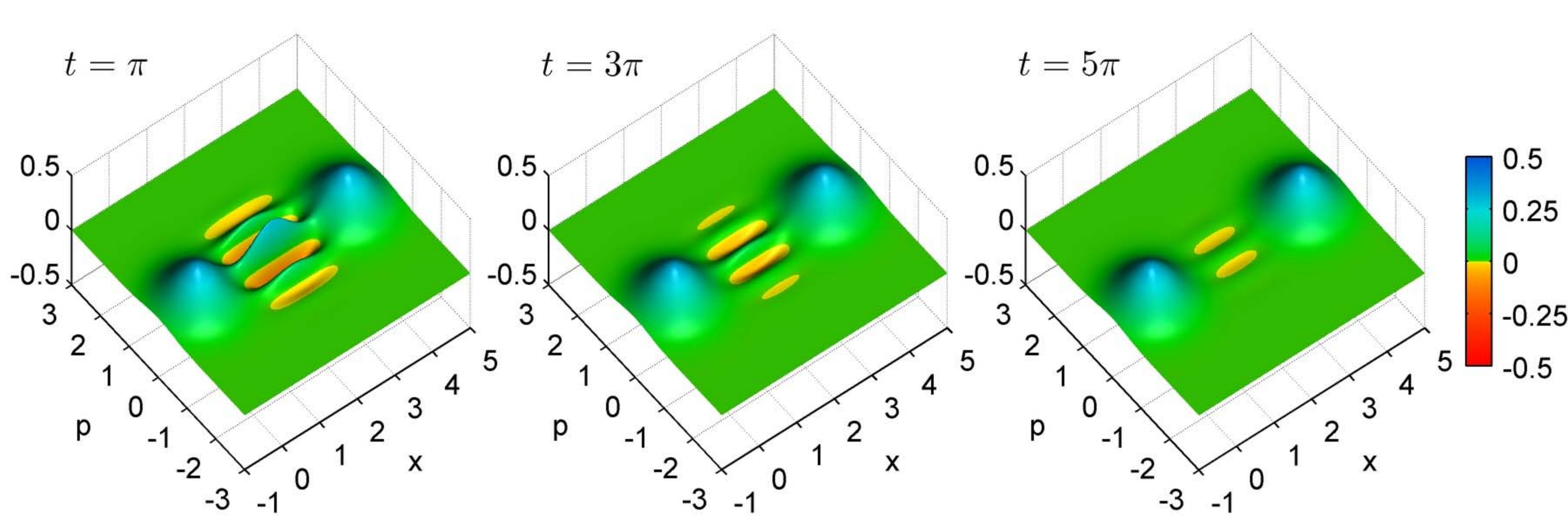}
	\end{center}
	\caption{
		Wigner function of the system in the presence of environmentally induced decoherence for $T_b = T_{EID}/64$, $\kappa = 2$ and $\hbar = \omega_c = m = 1$.
	}
	\label{fig:Wigner-Decoherence}
\end{figure}

Following Zurek \cite{Zurek2003RMP}, we note that in the macroscopic regime (to highest order in $\hbar^{-1}$), the master equation is dominated by the diffusion term proportional to $D/\hbar^2$.
Evaluating it in the position basis, one finds the time scale:
\be{eq-time}
	\tau_{\textrm{dec}} = \frac{\hbar^2}{D(\Delta x)^2} 	= \frac{\hbar Q}{16 k_B T_b \kappa^2},
\ee
where $\Delta x = \sqrt{8} \kappa x_0$, as before.
A calculation of the Wigner function which includes decoherence of the off-diagonal elements with the above dependence shows how the non-classicality of the state is dissipated with time (\fref{fig:Wigner-Decoherence}).

An exact open quantum system analysis of the experimental setup based on eqn.~\eref{eq-Master} has been performed by Bassi et al. \cite{Bassi2005PRL} and Bern\'{a}d et al. \cite{Bernad2006PRL}.
The former authors neglect the term proportional to $p$ in eqn.~\eref{eq-Master}  and solve the resulting equation for the off-diagonal matrix elements of the reduced photon density matrix.
The latter authors use the full equation.
The results for the decoherence of the revival peaks in those papers are remarkably close to the above estimate, both predicting a longer coherence time by only a factor of $8/3$.
The order of magnitude is thus well captured by \eref{eq-time}.

For an optomechanical system the important parameter is the mechanical quality factor, $Q$.
It is convenient to define a characteristic environmentally induced decoherence temperature:
\be{eq-TEID}
T_{EID} = \frac{\hbar \omega_c Q}{k_B}.
\ee
With this definition, the decoherence time \eref{eq-time} can be written as $ \tau_{\textrm{dec}}^{-1} = 16 \kappa^2 \omega_c \left(\frac{T_{b}}{T_{EID}}\right) $.
Above this temperature the interference revival peak will be drastically reduced in magnitude.
We note that the environmentally induced decoherence rate is dependent only on the bath temperature, $T_b$, not the effective temperature of the cantilever mode, $T$, which can be made different from the bath temperature by optical cooling (see \sref{sec-optical-cooling}).
Since a high-Q resonator is only weakly coupled to the bath, it is sufficient to treat these two temperatures as independent.

\subsection{Gravitationally Induced Quantum Collapse}
To explain the apparent classicality of the macroscopic world, it has been suggested that there may be a quantum state collapse mechanism for large objects, possibly induced by mass.
Several  proposals have been made which lead to such a collapse, among them reformulations of quantum mechanics \cite{Ghirardi1986PRD, Weinberg1989Annals} and the use of the intrinsic incompatibility between general relativity and quantum mechanics \cite{Karolyhazy1966, Penrose1996, Diosi1989PRA}.
Unlike environmentally induced decoherence, which is largely a nuisance in the realization of a massive superposition experiment, measurement of a mass induced collapse would be evidence of new physics and is hence of considerable interest.

Here we review the gravitational collapse model given by Penrose~\cite{Penrose1996}.
Penrose argues that a superposition of a massive object will result in a co-existence of two different space-time geometries which cannot be matched in a coordinate independent way.
Any difference in the causal structure will then generate different time translation operators $\partial / \partial t $ in the respective space-times.
Only an asymptotic identification would be possible, but if a local notion is required the failure to identify a single time structure for two superposed space-times will be a fundamental obstacle to unitary quantum evolution. Any time translation operator $\partial / \partial t $ in such a superposition of space-times will have an intrinsic error and hence a unitary evolution cannot take place indefinitely.
This will eventually result in a collapse of the superposed state.

To give an order of magnitude estimate for the identification of the two superposed space-times, Penrose uses the Newtonian limit of gravity including the principle of general covariance.
The error is quantified by the difference of free falls (geodesics) throughout both space-times, which turns out to correspond to the gravitational self energy $\Delta E $ of the superposed system, defined the following way:
\bems{eq-self-energy}
E_{i, j}&=& - G \int\!\!\! \int\!\! d\vec{r_1} d\vec{r_2} \, \frac{\rho_i(\vec{r_1}) \rho_j(\vec{r_2})}{\left|\vec{r_1} - \vec{r_2} \right|}
\nems{eq-self-energy-2}
\Delta E&=& 2 E_{1, 2} - E_{1, 1} - E_{2, 2},
\eems
where $\rho_1$ and $\rho_2$ are the mass distributions for the two states in question.
A similar result was obtained by Diosi ~\cite{Diosi1989PRA}.
This energy yields a timescale for the decay of a superposition, estimated by $\tau_G \approx \hbar / \Delta E $.

When attempting to apply this to the proposed superposition experiment, it is unclear precisely what form the mass distributions should take (see also \cite{Diosi2007JPA}).
For simplicity we will consider the mass to be evenly distributed over a number of spheres, corresponding to atomic nuclei, each with mass $m_1$, radius $a$, and the superposition states to be separated by a distance $\Delta x$.
The total mass is given by $m$, as before.
If the atomic spacing is much larger than the effective mass radius, the energy due to the interaction between different atomic sites is negligible and the gravitational self-energy is given by:
\be{eq-DE-spheres}
\Delta E = 2 G m m_1 \left( \frac{6}{5 a} - \frac{1}{\Delta x} \right) \quad \textrm{(given: } \Delta x \geq 2 a \textrm{)}.
\ee
If we set the sphere radius to be the approximate size of a nucleus ($a = 10^{-15}$ m) and use the parameters of an ideal optomechanical device ($m = 10^{-12}$ kg, $\omega_c = 2 \pi \times 1$ kHz, $\kappa = 1/\sqrt{2}$ and $m_1 = 4.7 \times 10^{-26}$ kg, the silicon nuclear mass), this results in a timescale of order milliseconds.
Alternatively, one could argue that the effective diameter of the spheres should be the ground-state wavepacket size ($a = x_0/2$).
With the maximum separation of the states ($\Delta x = \sqrt{8} \kappa x_0$), the resulting energy is:
\be{eq-DE-cantilever}
\Delta E = \frac{G m m_1}{x_0} \left( \frac{24}{5} - \frac{1}{\sqrt{2} \kappa} \right).
\ee
Using the ideal device parameters results in a timescale on the order of 1 second.

In order to practically measure such a collapse mechanism, we require the timescale to be not much larger than a mechanical period so that a significant visibility reduction is present in the first revival peak.
This means it may be possible to measure a mass-induced collapse effect with the proposed experiment, although we note that the collapse timescale given above is intended only to be a rough estimate.
To contrast with previous large superposition experiments, the collapse timescale for interferometry of large molecules like C$_{60}$~\cite{Arndt1999} is calculated to be $10^{10}$ s (using the nuclear radius, $a = 10^{-15}$ m, and assuming comparatively larger separation).
Other demonstrated experiments have similar or larger timescales, meaning a collapse mechanism of this type would have certainly been undetectable in all experiments to date.
\section{Prospects for Experimental Realization}
\label{section-experiment}
\subsection{Optomechanical Devices}
In practice, the experimental realization of a macroscopic quantum superposition is severely technically demanding.
Perhaps the most challenging aspect is achieving sufficient optical quality, which is required to put the cantilever into a distinguishable state via interaction with a single photon, i.e. $\kappa \gtrsim 1/\sqrt{2}$.
Although $\kappa$ can be increased by shortening the optical cavity, this will also reduce the ring-down time, making it extremely unlikely to observe a photon in the revival period.
A reasonable compromise is reached by requiring the optical finesse, $F$, be equal to the required number of round trips per period as given by \eref{eq-Kappa-2}.
In this case the fraction of photons still in the optical cavity after a mechanical period is $e^{-2 \pi}$ (.2\%), a small number but enough to measure the visibility on the timescale of hours.
This resulting requirement for the finesse has a rather intuitive form:
\be{eq-Finesse}
F \gtrsim \frac{\lambda}{2 x_0}.
\ee

\begin{figure}[!htp]
	\begin{center}
		\includegraphics{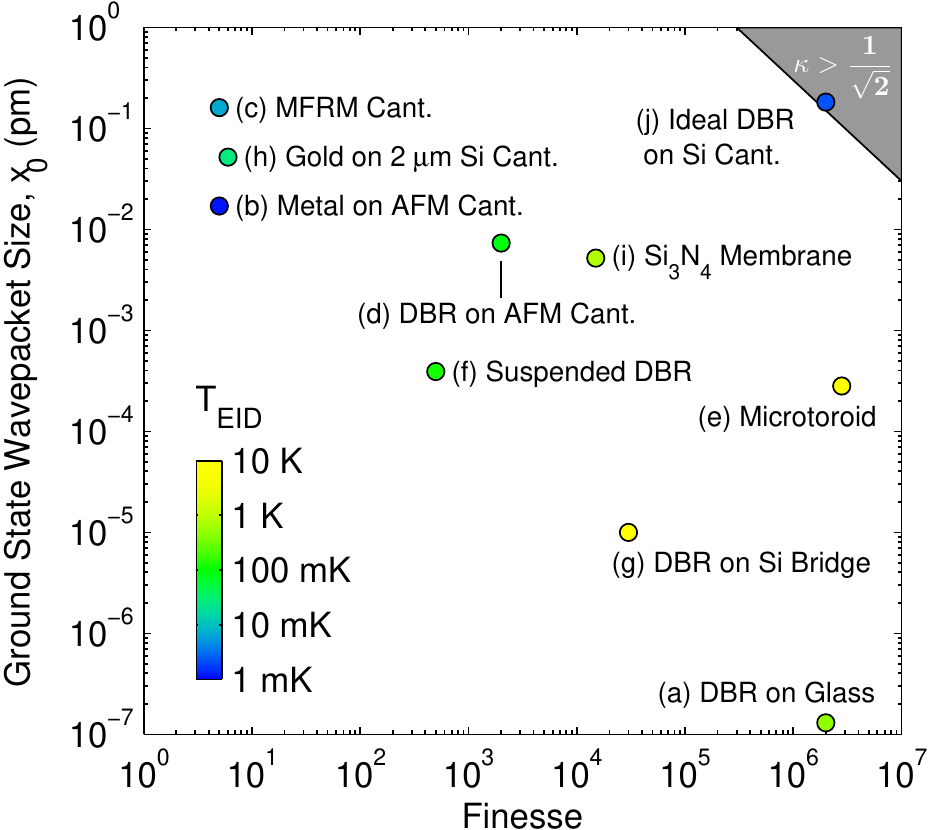}
	\end{center}
	\caption{	A comparison of opto-mechanical devices, showing the finesse and size of the ground state wavepacket, $x_0 = \sqrt{\hbar/m \omega_c}$.
	All points apart from (j) are based on experimental results.
	The shaded area in the upper right corresponds to $\kappa = 1/\sqrt{2}$ for visible light ($\lambda = 600$ nm).
	The color of each point corresponds to the characteristic environmentally induced decoherence temperature, $T_{EID} = \hbar \omega_c Q / k_b$.
	Many of the devices are the subject of ongoing research, and so the listed parameters should be regarded as approximate. \\
	\begin{tabular}{r p{4.5in}}
		\textbf{(a)}&A dielectric Bragg reflector (DBR) with $F = 2 \times 10^6$ deposited on a cm size mirror.  \\
		\textbf{(b)}&Metal deposited on a conventional atomic force microscopy (AFM) cantilever (for example, \cite{Metzger2004}).\\
		\textbf{(c)}&A thin silicon cantilever used in magnetic force resonance microscopy (MFRM)~\cite{Mamin2001}.\\
		\textbf{(d)}&A Focused Ion Beam milled DBR mirror glued to a commercial AFM cantilever~\cite{Kleckner2006PRL}.\\
		\textbf{(e)}&Microtoroidal resonator~\cite{Schliesser2007}.  ($\kappa$ is not given by \eref{eq-Kappa-2} because of a different geometry.)\\
				\textbf{(f)}&Resonator made of a suspended DBR bridge~\cite{Gigan2006Nature}.\\
		\textbf{(g)}&DBR deposited on a silicon bridge resonator~\cite{Arcizet2006Nature}.\\
		\textbf{(h)}&A 2 $\mu$m silicon resonator with gold deposited on it~\cite{Favero2007APL}.\\
		\textbf{(i)}&Commercial Si$_3$N$_4$ membrane in a high finesse optical cavity~\cite{Zwickl2008}.\\
		\textbf{(j)}&Theoretical device with a tiny, high finesse DBR mirror attached to a cantilever similar to those used in MFRM experiments ($m = 10^{-12}$ kg, $\omega_c = 2 \pi \times 500$ Hz, $F = 2 \times 10^6$)\\
	\end{tabular}
	}
	\label{fig-Devices}
\end{figure}

In order to prevent diffraction from limiting the finesse, the mirror on the cantilever needs a diameter of order 10 microns or larger \cite{Kleckner2006PRL}.
If the mirror is a dielectric Bragg reflector, the conventional choice for achieving very high optical quality, the required finesse is of order $10^6 - 10^7$ given the minimum resulting mass and assuming it is placed on a cantilever with frequency $\sim$ 1 kHz.
Finesses of over $10^6$ have been realized in several experiments with larger, cm size, dielectric mirrors (for example, \cite{Rempe92}), so the primary challenge in using these mirrors is finding a way to micro-fabricate them without degrading their properties.
State of the art is currently $F = 10^4-10^5$, although rapid progress has been made in recent years due to a growing interest in optomechanical systems in general. See \fref{fig-Devices} for a comparison of different devices.
An interesting alternative to the tiny mirror on the cantilever approach is the so called ``membrane in the middle''.
In this case the optomechanical element is a dielectric membrane placed between two high quality mirrors; the cavity detuning induced by motion of the membrane produces a result functionally equivalent to moving an end mirror on a mechanical resonator.
Commercially available silicon nitride membranes have recently been demonstrated in cavities with finesses of over $10^4$ and with remarkably high mechanical quality factors, $Q > 10^7$~\cite{Zwickl2008}.
In theory, this type of system would require a lower finesse to achieve a superposition, as the thickness of the optical element can be an order of magnitude less than a dielectric mirror.
To take advantage of this, however, would require the membranes be micro-fabricated into cantilever or bridge-resonator structures to reduce their total mass, something that has not yet been attempted.

The other important parameter for an optomechanical system is the mechanical quality factor, $Q$, governing the characteristic environmentally induced decoherence temperature $T_{EID}$, as defined in \eref{eq-TEID}.
Optomechanical devices have already been demonstrated for which $T_{EID}$ is experimentally accessible with common cryogenic techniques (\fref{fig-Devices}), although operating the devices in the sub-Kelvin regime is likely to be difficult.
Resonators used in magnetic force resonance microscopy experiments, which have similar mechanical properties, have been cooled to temperatures of around 100 mK, limited by heating due to optical absorption in the readout~\cite{Mamin2001}.
Although the magnitude of this effect should be smaller for high finesse optomechanical systems due to lower absorption and incident light levels, at temperatures of order 1 mK absorption of even single photons should produce non-negligible heating~\cite{Kleckner2006Nature}.

\subsection{Optical Cooling}
\label{sec-optical-cooling}
As stated above, unambiguous observation of a macroscopic quantum superposition is possible only when the cantilever's fundamental mode is in a low phonon quantum number state.
Given that this requires temperatures of less than 1 $\mu$K for kHz resonators, the only way to practically obtain this is optical feedback cooling.
There are two primary forms of optical feedback cooling, referred to as ``active'' and ``passive''.
Active feedback cooling uses the optical cavity to read out the position of the cantilever, and then an electronic feedback loop creates a force on the cantilever (using, e.g., a second intensity modulated laser) to dampen the motion of its fundamental mode.
Because the effective damping force is not subjected to thermal fluctuations, this is equivalent to coupling the system to a zero temperature thermal bath, and so the effective temperature of the fundamental mode can be dramatically reduced.
Passive feedback cooling uses the finite ring-down time of the optical cavity to intrinsically produce a similar damping force without the use of an external feedback loop.
Note that neither type of cooling significantly reduces the temperature of the environmental bath, so the environmentally induced decoherence timescale is virtually unaffected by optical cooling.
Both active~\cite{Kleckner2006Nature, Cohadon1999prl, Poggio2007} and passive~\cite{Thompson2008, Schliesser2007, Gigan2006Nature, Arcizet2006Nature, Favero2007APL, Schliesser2006, Corbitt2007prl, Groblacher2007, Corbitt2007arxiv} feedback cooling have been experimentally demonstrated by many groups, in some cases achieving cooling factors of well over 10$^3$.

\begin{figure}
\begin{center}
\includegraphics{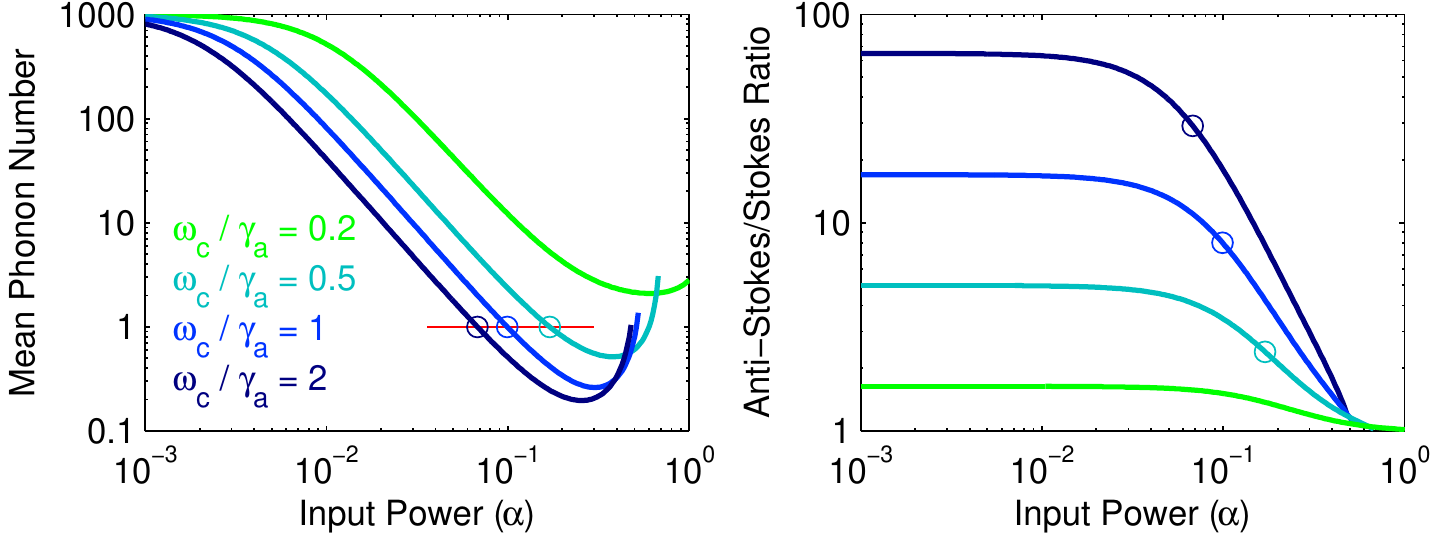}
\end{center}
\caption{
	Left: Mean phonon number, $\bar{n}$ as a function of power for passive optical feedback cooling.  Right: Anti-Stokes/Stokes ratio.
	The theoretical model is derived from Marquardt et al.~\cite{Marquardt2007PRL}.
	The input optical field strength is given in terms of a dimension less power, $\alpha = \sqrt{2 \bar{n}_a} \kappa$ where $\bar{n}_a$ is the mean number of photons in the optical cavity.
	$\gamma_a$ is the power decay constant for the optical cavity.
	Pump photons are detuned from the cavity resonance by $\Delta = -\omega_c$.
	When $\bar{n} = 1$, the Anti-Stokes/Stokes ratio decreases to half its low field ($\alpha \to 0$) limit, shown with circles in (b).
	The ratio, which can be measured in the light leaving the cavity, provides a direct method to determine the effective temperature of the cantilever.
}
\label{fig-Cooling}
\end{figure}

If one operates below the environmentally induced decoherence temperature given above, it is theoretically possible to cool the fundamental mode of the cantilever near the ground state using either active~\cite{Courty2001} or passive optical feedback cooling~\cite{WilsonRae2007, Marquardt2007PRL, Bhattacharya2007PRL}, although this has yet to be demonstrated experimentally.
Although heating due to optical absorption and linewidth of the drive laser are serious concerns ~\cite{Diosi2000PRA}, these do not present fundamental obstacles.
In the limit that the ring-down time is comparable to the mechanical period, as indeed it must be for observing a macroscopic superposition, passive cooling should be more effective.
The equilibrium phonon occupation number of the cantilever as a function of pumping power is shown in \fref{fig-Cooling}; the situation where $N = F$, as discussed above, corresponds to $\omega_c/\gamma_a = 1$.
Conveniently, passive cooling also provides a method to directly measure the phonon number of the cantilever by measuring the ratio of anti-Stokes to Stokes shifted photons in the outgoing cavity field (see also \fref{fig-Cooling})~\cite{WilsonRae2007, Marquardt2007PRL}.
In the limit of low pumping power and minimal cooling, this ratio remains constant, but begins to rapidly decrease when the ground state is approached.
When the ratio is less than half the low power value, the mean phonon number, $\bar{n}$, is less than one, providing a clear indication of ground state cooling.
Because this type of cooling can be easily integrated with the proposed macroscopic superposition experiment, it presents an ideal method for putting the system in a known low phonon number state.

\section{Conclusion}
A detailed analysis of the effects of finite temperature on the proposed massive superposition experiments show that a fully unambiguous demonstration requires low fundamental mode temperatures, $\bar{n} \lesssim 1$.
Despite this, observation of a revival of the interference visibility can be used to strongly imply the existence of a superposition at higher temperatures, as proposed in \cite{Marshall2003PRL}.
Additionally, the magnitude of the visibility revival provides an opportunity to test environmentally induced decoherence models and possibly measure proposed mass-induced collapse mechanisms.
Although such an experiment is difficult to realize, comparison to several related experiments suggests it should be technologically feasible.
This is greatly aided by growing interest in developing high quality micro-optomechanical devices for a range of applications.
Additionally, recently developed optical feedback cooling techniques can be used to obtain fundamental mode temperatures far lower than are conventionally accessible, possibly even cooling to the ground state.

\section{Acknowledgments}
The authors would like to thank C. Simon and L. Diosi for useful discussions.
I. P. thanks J. Bosse for support.
This work was supported in part by the National Science Foundation (grants PHY-0504825 and PHY05-51164), Marie-Curie EXT-CT-2006-042580 and the Stichting voor Fundamenteel Onderzoek der Materie (FOM).
\section{References}
\bibliography{refs2}

\end{document}